\documentclass[fleqn]{llncs}

\usepackage{times}
\usepackage[utf8]{inputenc}
\usepackage[T1]{fontenc}
\usepackage{url}

\usepackage{hyperref}
\usepackage{amsmath}
\usepackage{amssymb}

\usepackage{listings}
\lstnewenvironment{curry}[1][]{\lstset{#1}}{}
\lstnewenvironment{prolog}{}{}
\newcommand{\listline}{\vrule width0pt depth1.25ex}
\renewcommand{\tt}{\ttfamily}
\newcommand{\codefont}{\small\tt}
\newcommand{\code}[1]{\mbox{\codefont{#1}}} % program text in normal text
\newcommand{\ccode}[1]{``\code{#1}''}  % cited program text in normal text
\newcommand{\fcode}[1]{\mbox{\codefont{\footnotesize{#1}}}} % code in footnote
\newcommand{\us}{\char95} % underscore
\begin{document}

\title{CurryCheck: Checking Properties of Curry Programs}
\author{Michael Hanus}
\institute{
Institut f\"ur Informatik, CAU Kiel, D-24098 Kiel, Germany \\
\email{mh@informatik.uni-kiel.de}
}

\maketitle

\vspace*{-3ex}
\begin{abstract}
We present CurryCheck, a tool to automate the testing of
programs written in the functional logic programming language Curry.
CurryCheck executes unit tests as well as
property tests which are parameterized over one or more arguments.
In the latter case, CurryCheck tests these properties
by systematically enumerating test cases so that, for smaller finite domains,
CurryCheck can actually prove properties.
Unit tests and properties can be defined in a Curry module
without being exported.
Thus, they are also useful to document the intended semantics
of the source code.
Furthermore, CurryCheck also supports the automated checking
of specifications and contracts occurring in source programs.
Hence, CurryCheck is a useful tool that contributes
to the property- and specification-based development
of reliable and well tested declarative programs.
\end{abstract}

%%%%%%%%%%%%%%%%%%%%%%%%%%%%%%%%%%%%%%%%%%%%%%%%%%%%%%%%%%%%%%%%%%%%%%%
\section{Motivation}
\label{sec:motivation}

Testing is an important step to get confidence
in the functionality of a program.
The advantage of testing compared to program verification
is its potential for automation.
If we do not execute test cases only manually for some inputs
but encode them as input to test frameworks,
we can automatically run and repeat them when the software
is further developed, which is also known as regression testing.

A difficulty in testing is to find appropriate inputs for the
individual tests.
For this purpose, property-based testing has been proposed,
well known in the functional language Haskell with the
QuickCheck tool \cite{ClaessenHughes00}.
Basically, properties are predicates parameterized over one
or more arguments.
QuickCheck automates the test execution by applying properties
to randomly generated test inputs.
Since this idea is particularly reasonable for declarative
languages, it is been adapted in different forms
to functional and logic programming languages.
For instance, SmallCheck \cite{RuncimanNaylorLindblad08}
and GAST \cite{KoopmanAlimarineTretmansPlasmeijer03}
focus on a systematic enumeration of test inputs
for functional programs, 
PropEr \cite{PapadakisSagonas11} adapts ideas of QuickCheck
to the concurrent functional language Erlang,
PrologCheck \cite{AmaralFloridoSantosCosta14}
transfers and extends ideas of QuickCheck to Prolog,
and EasyCheck \cite{ChristiansenFischer08FLOPS}
exploits functional logic programming features
to property-based testing of Curry programs.

CurryCheck follows the same ideas.
Actually, it is based on EasyCheck to define properties.
However, CurryCheck is intended as a comprehensive tool to simplify
the automation of test execution.
To use CurryCheck, properties are interspersed
into the program as top-level definitions.
Thus, properties are used to document the intended
semantics of the source code, which also supports
test-driven program development known as ``extreme programming.''
When CurryCheck is applied to a (set of) Curry modules,
it extracts all properties, generates a program to test these
properties, executes this generated program, and reports any errors.
Furthermore, CurryCheck also analyzes possible contracts
\cite{AntoyHanus12PADL} provided in source programs
and generates properties to test these contracts.
Thanks to this automation, CurryCheck is a useful tool
for continuous integration and deployment processes.
Actually, it is used for this purpose in the Curry implementations
PAKCS \cite{Hanus16PAKCS} and KiCS2 \cite{BrasselHanusPeemoellerReck11}.

In this paper we present the ideas and usage of CurryCheck.
After a review of the main features of Curry in the next section,
we introduce properties in Sect.~\ref{sec:properties}
and explain how they are tested in Sect.~\ref{sec:testproperties}.
The support of CurryCheck to define test inputs is presented
in Sect.~\ref{sec:gentestdata}.
CurryCheck's support for contract checking is described in
Sect.~\ref{sec:contracts}.
Some initial features of CurryCheck to combine testing and verification
are sketched in Sect.~\ref{sec:verify}.
We report about our practical experience with CurryCheck
in Sect.~\ref{sec:impl}
before we compare CurryCheck to some related tools and conclude.

%%%%%%%%%%%%%%%%%%%%%%%%%%%%%%%%%%%%%%%%%%%%%%%%%%%%%%%%%%%%%%%%%%%%%%%%%%%%
\section{Functional Logic Programming and Curry}
\label{sec:flp}

Functional logic languages \cite{AntoyHanus10CACM,Hanus13}
integrate the most important features
of functional and logic languages in order to provide a variety
of programming concepts.
They support functional concepts like
higher-order functions and lazy evaluation
as well as logic programming concepts like
non-deterministic search and computing with partial information.
This combination led to new design patterns \cite{AntoyHanus11WFLP}
as well as better abstractions for application programming.
The declarative multi-paradigm language Curry \cite{Hanus97POPL}
is a modern functional logic language with advanced programming concepts.
In the following, we briefly review some features of Curry
relevant for this paper.
More details can be found in recent surveys on
functional logic programming \cite{AntoyHanus10CACM,Hanus13}
and in the language report \cite{Hanus16Curry}.

The syntax of Curry is close to Haskell \cite{PeytonJones03Haskell}.
In addition to Haskell, Curry allows \emph{free} (\emph{logic}) \emph{variables}
in rules and initial expressions.
Function calls with free variables are evaluated by a possibly
non-deterministic instantiation of demanded arguments.

\begin{example}\label{ex:concdup}
The following simple program shows the functional and logic features
of Curry. It defines the well-known list concatenation
and an operation that returns
some element of a list having at least two occurrences:
\begin{curry}
(++) :: [a] -> [a] -> [a]        someDup :: [a] -> a
[]     ++ ys = ys                  someDup xs | xs == _$\,$++$\,$[x]$\,$++$\,$_$\,$++$\,$[x]$\,$++$\,$_
(x:xs) ++ ys = x : (xs ++ ys)                 = x    where$\;$x$\;$free
\end{curry}
The (optional) type declaration (\ccode{::}) of the operation \ccode{++}
specifies that \ccode{++} takes two lists as input and produces
an output list, where all list elements are of the same (unspecified) type.
Since \ccode{++} can be called with free variables in arguments,
the condition in the rule of \code{someDup}
is solved by instantiating \code{x} and
the anonymous free variables \ccode{\us} to appropriate values
before reducing the function calls.
This corresponds to narrowing \cite{Slagle74,Reddy85},
but Curry narrows with possibly non-most-general unifiers
to ensure the optimality of computations \cite{AntoyEchahedHanus00JACM}.
\end{example}
Note that \code{someDup} is a \emph{non-deterministic operation}
since it might deliver more than one result for a given argument,
e.g., the evaluation of \code{someDup$\,$[1,2,2,1]} yields the values
\code{1} and \code{2}.
Non-deterministic operations, which can formally be
interpreted as mappings from values into sets of values \cite{GonzalezEtAl99},
are an important feature
of contemporary functional logic languages.
Hence, Curry has also a predefined \emph{choice} operation:
\begin{curry}
x ? _  =  x
_ ? y  =  y
\end{curry}
Thus, the expression \ccode{0$~$?$~$1} evaluates to \code{0} and \code{1}
with the value non-deterministically chosen.

\emph{Functional patterns} \cite{AntoyHanus05LOPSTR} are useful
to define some operations more easily.
A functional pattern is a pattern occurring in an argument
of the left-hand side of a rule containing defined operations
(and not only data constructors and variables).
Such a pattern abbreviates the set of all standard patterns to which the
functional pattern can be evaluated (by narrowing).
For instance, we can rewrite the definition of \code{someDup} as
\begin{curry}
someDup (_++[x]++_++[x]++_) = x
\end{curry}
Functional patterns are a powerful feature to express arbitrary selections
in tree structures, e.g., in XML documents \cite{Hanus11ICLP}.
Details about their semantics and a constructive implementation
of functional patterns by a demand-driven unification procedure
can be found in \cite{AntoyHanus05LOPSTR}.

Curry has also features which are useful for application programming,
like \emph{set functions}  \cite{AntoyHanus09} to encapsulate
non-deterministic computations,
\emph{default rules} \cite{AntoyHanus16PADL} to deal
with partially specified operations and negation,
and standard features from functional programming,
like modules or monadic  I/O \cite{Wadler97}.
Other features are explained when they are used in the following.

%%%%%%%%%%%%%%%%%%%%%%%%%%%%%%%%%%%%%%%%%%%%%%%%%%%%%%%%%%%%%%%%%%%%%%%%%
\section{Properties}
\label{sec:properties}

In this section we briefly discuss which kind of program properties
to be tested are supported by CurryCheck.
Since CurryCheck extends the functionality of
EasyCheck \cite{ChristiansenFischer08FLOPS},
it supports all kinds of EasyCheck's properties
which we review first.

Properties are defined top-level entities with a distinct type (see below).
Thus, their syntax and type-correctness
can be checked by the standard front end of any Curry system.
Properties do not require a specific naming convention
but CurryCheck recognizes them by their type.
Moreover, the name and position of the property in the source file
are used by CurryCheck to identify properties when errors are reported.

For instance, consider the list concatenation operation \ccode{++}
defined in Example~\ref{ex:concdup}.
Before discussing general properties, we define some
unit tests for fixed arguments, like
\begin{curry}
concNull12   =  []   ++ [1,2] -=- [1,2]
concCurry    =  "Cu" ++ "rry" -=- "Curry"
\end{curry}
The infix operator \ccode{-=-} specifies a test which is
successful if both sides have single values which are identical
(we will later see tests for non-deterministic operations).
Since the expressions can be of any type (of course, the two
arguments must be of the same type),
the operator is polymorphic and has the type
\begin{curry}
(-=-) :: a -> a -> Prop
\end{curry}
Hence, all entities defined above have type \code{Prop}.

The power of CurryCheck and similar property-based test frameworks
comes from the fact that we can also test properties which are
parameterized over some input data.
For instance, we can check whether the concatenation operation
is associative by:
\begin{curry}
concIsAssociative xs ys zs = (xs++ys)++zs -=- xs++(ys++zs)
\end{curry}
This property is parameterized over three input values
\code{xs}, \code{ys}, and \code{zs}.
To test this property, CurryCheck guesses values for these parameters
(see below for more details) and tests the property for these values:
\begin{curry}
concIsAssociative_ON_BASETYPE (module ConcDup, line 18):
 OK, passed 100 tests.
\end{curry}
Indicated by the suffix \code{\us{}ON\us{}BASETYPE},
we see another feature of CurryCheck.
If properties are polymorphic in their input values
(the above property has type
\code{[a]$\;\to\;$[a]$\;\to\;$[a]$\;\to\;$Prop}),
CurryCheck specializes the type to some base type,
since there is no concrete value of a polymorphic type
(and EasyCheck would fail on such properties).
As a default, CurryCheck uses the predefined type \code{Ordering}
having the three values \code{LT}, \code{EQ}, \code{GT}
(another more involved method to
instantiate polymorphic types in purely functional programs
can be found in \cite{BernardyJanssonClaessen10}).
This default type can be changed to other
base types, like \code{Bool}, \code{Int}, or \code{Char},
with a command-line option.
One could also provide an explicit type declaration for the
property. For instance, we can test the commutativity
of the list concatenation on lists of integers by the property
\begin{curry}
concIsCommutative :: [Int] -> [Int] -> Prop
concIsCommutative xs ys = (xs ++ ys) -=- (ys ++ xs)
\end{curry}
Of course, this property does not hold so that CurryCheck
reports an error together with a counter-example:
\begin{curry}
$\ldots$
concIsCommutative (module ConcDup, line 20) failed
Falsified by 8th test.
Arguments: [-1] [-3]
Results:   ([-1,-3],[-3,-1])
\end{curry}
Note that the arguments of a test are ordinary expressions
so that one can use any defined operation in the tests.
For instance, we can (sucessfully) check whether the list concatenation
is the addition on their lengths:
\begin{curry}
concAddLengths xs ys = length xs + length ys -=- length (xs++ys)
\end{curry}
Since Curry covers also logic programming features,
CurryCheck supports the testing of non-deterministic properties.
For instance, one can check whether an expression reduces to some
given value with the operator is \ccode{\char126>}:
\begin{curry}
someDup1 = someDup [1,2,1,2] ~> 1
\end{curry}
Another important operator is \ccode{<\char126>} which
denotes a test which succeeds if both arguments
have the same set of values.
We can write unit tests by enumerating all expected values
with the choice operator \ccode{?}:
\begin{curry}
someDup12 = someDup [1,2,1,2,1] <~> (1$\,$?$\,$2)
\end{curry}
It should be noted that the operator \ccode{<\char126>}
really compares sets and not multi-sets:
Although the evaluation of \code{someDup$\;$[1,2,1,2,1]}
returns the value \code{1} three times in a typical Curry system,
the property \code{someDup12} holds.
This is intended since CurryCheck tests declarative properties which 
are independent of specific compiler optimizations
(this is in contrast to PrologCheck which tests operational
properties like multiplicity of answers and modes
\cite{AmaralFloridoSantosCosta14}).

As another example, consider the following definition
of a permutation of a list by exploiting a functional pattern
to select some element in the argument list:
\begin{curry}
perm (xs++[x]++ys) = x : perm (xs++ys)
perm []            = []
\end{curry}
An important property of a permutation is that the length of
the list is not changed. Hence, we check it by the property
\begin{curry}
permLength xs = length (perm xs) <~> length xs
\end{curry}
Note that the use of \ccode{<\char126>} (instead of \ccode{-=-})
is relevant since non-deterministic values are compared.
Actually, the left argument evaluates to many (identical) values.

We might also want to check whether our definition of \code{perm} computes the
correct number of solutions. Since we know that a list of length $n$
has $n!$ permutations, we write the following property,
where \code{fac} is the factorial function and
the property \code{$x$ \# $n$} is satisfied if $x$ has $n$ different values:
\begin{curry}
permCount :: [Int] -> Prop
permCount xs = perm xs # fac (length xs)
\end{curry}
However, this test will be falsified with the test input \code{[1,1]},
since this list has only one permuted value (actually, both computed
values are identical).
We can obtain a correct property if we add the condition
that all elements in the input list \code{xs} are pairwise different.
For this purpose, we use a \emph{conditional property}:
the property \code{$b$ ==> $p$} is satisfied if $p$
is satisfied for all values where $b$ evaluates to \code{True}.
If the predicate \code{allDifferent} is satisfied iff its argument list
does not contain duplicated elements,
then we can reformulate our property as follows:
\begin{curry}
permCount xs = allDifferent xs ==> perm xs # fac (length xs)
\end{curry}
Furthermore, we want to check the existence of distinguished permutations.
For this purpose, consider a predicate to check whether a list is sorted:
\begin{curry}
sorted :: [Int] -> Bool
sorted []       = True
sorted [_]      = True
sorted (x:y:zs) = x<=y && sorted (y:zs)
\end{curry}
Then we can check whether there are sorted permutations
(the property \ccode{eventually$\;x$} is satisfied if some value
of $x$ is \code{True}):
\begin{curry}
permIsEventuallySorted :: [Int] -> Prop
permIsEventuallySorted xs = eventually (sorted (perm xs))
\end{curry}
Property-based testing is appropriate for declarative languages
since the absence of side effects allows the execution
of tests on any number of test data without influencing
the individual tests.
Nevertheless, real programming languages have to deal with the
real world so that they support also I/O operations.
Clearly, such operations should also be tested.
Although there are methods to test monadic code \cite{ClaessenHughes02},
Curry supports only I/O monadic operations where testing with
arbitrary data seems not reasonable.
Therefore, CurryCheck supports only non-parameterized
unit tests for I/O operations.
For instance, the test \code{($a$ `returns` $x$)} is satisfied
if the I/O action $a$ returns the value $x$.
For instance, we can test whether writing a file and reading it
yields the same contents:
\begin{curry}
writeReadFile = (writeFile "TEST" "Hello" >> readFile "TEST")
                  `returns` "Hello"
\end{curry}
Since CurryCheck executes the tests written in a source program
in their textual order, one can write also several I/O tests
whose side effects depend on each other.
For instance, we can split the previous I/O test into two
consecutive tests:
\begin{curry}
writeTestFile = (writeFile "TEST" "Hello") `returns` ()
readTestFile  = (readFile "TEST")          `returns` "Hello"
\end{curry}

%%%%%%%%%%%%%%%%%%%%%%%%%%%%%%%%%%%%%%%%%%%%%%%%%%%%%%%%%%%%%%%%%%%%%%%%%
\section{Testing Properties}
\label{sec:testproperties}

After having seen several methods to define properties,
we sketch in this section how they are actually tested.
Our motivation for the development of CurryCheck is manifold:
\begin{enumerate}
\item
Properties are an executable documentation for the
intended semantics of operations.
\item
Properties increase the confidence in the quality of the developed
software.
\item
Properties can be used for software verification
by proving their validity for all possible input data.
\end{enumerate}
The first point is supported by interspersing properties
into the source code of the program instead of putting them
into separate files.
Thus, properties play the same role as comments or type annotations:
they document the intended semantics.
Hence, they can be extracted and put into the program
documentation by automatic documentation tools
\cite{Hanus02WFLP,Hermenegildo00}.
In order to avoid that properties influence the interface
of a module, they do not need to be exported.
As an example, consider the following simple module defining
the classical list reverse operation (the imported module
\code{Test.EasyCheck} contains the definitions of the property combinators
introduced in Sect.~\ref{sec:properties}):
\begin{curry}
module Rev(rev) where$\listline$
import Test.EasyCheck$\listline$
rev :: [a] -> [a]
rev []     = []
rev (x:xs) = rev xs ++ [x]$\listline$
revLength  xs = length (rev xs) -=- length xs
revRevIsId xs = rev (rev xs) -=- xs
\end{curry}
We can run all tests of this module by invoking the CurryCheck executable
with the name of the module:\footnote{One can also provide several
module names so that they are tested at once.
Furthermore, CurryCheck has various options to influence
the number of test cases, default types for polymorphic properties, etc.}
\begin{curry}
> currycheck Rev
Analyzing module 'Rev'...
...
Executing all tests...
revLength_ON_BASETYPE (module Rev, line 9):
 OK, passed 100 tests.
revRevIsId_ON_BASETYPE (module Rev, line 10):
 OK, passed 100 tests.
\end{curry}
Although module \code{Rev} only exports the operation \code{rev},
all properties defined in the top-level of \code{Rev}
are passed to the underlying EasyCheck library for testing.
For this purpose, CurryCheck creates a copy of this module
where all entities are exported (note that the original module
cannot be modified since it might be imported to other modules
to be tested).
For each property a corresponding call to
an operation of EasyCheck is generated which actually
performs the generation of test data, runs the test,
and collects all results which are passed back to CurryCheck.
Furthermore, polymorphic properties are not checked
but a corresponding new property on the default base type
is generated which calls the polymorphic property.
For instance, if the default base type is \code{Int},
CurryCheck generates the new property
\begin{curry}
revRevIsId_ON_BASETYPE :: [Int] -> Prop
revRevIsId_ON_BASETYPE = revRevIsId
\end{curry}
which is actually checked instead of \code{revRevIsId}.
Note that it might lead to a failure if the type of \code{revRevIsId}
is directly specialized, since the polymorphic property  \code{revRevIsId}
might be used in other property definitions with a different specialized type.

After these preparations, EasyCheck tests the properties
by generating test data as described in
\cite{ChristiansenFischer08FLOPS}.
EasyCheck does not use random generators
like QuickCheck or PrologCheck,
but it exploits functional logic programming features
to enumerate possible input values.
Since logic variables are equivalent to non-deterministic
generators \cite{AntoyHanus06ICLP},
one can evaluate a logic variable of a particular type
in order to get all values of this type in a non-deterministic manner.
For instance, if we evaluate the Boolean variable
\code{b::Bool}, we obtain the values \code{False} and \code{True}
as results. Similarly, for \code{bs::[Bool]}
we obtain values like \code{[]}, \code{[False]}, \code{[True]},
\code{[False,False]}, etc.
In order to select a finite amount of these infinite values,
one can use Curry's feature for encapsulated search to
collect all non-deterministic results in a tree structure,
traverse the tree with some strategy and return the
result of the traversal into a list.
If one selects only a finite amount of this list,
the lazy evaluation strategy of Curry ensures a finite computation
even if the tree is infinite.
Based on these features, the EasyCheck library contains an operation
\label{sec:valuesOf}
\begin{curry}
valuesOf :: a -> [a]
\end{curry}
which computes the list of all values of the given argument
according to a fixed strategy (in the current implementation by
randomized level diagonalization \cite{ChristiansenFischer08FLOPS}).
Hence, we can get 20 values for a list of integers by
\begin{curry}
$\ldots$> take 20 (valuesOf (_::[Int]))
[[],[-1],[-3],[0],[1],[-1,0],[-2],[0,0],[3],[-1,1],[-3,0],[0,1],
[2],[-1,-1],[-5],[0,-1],[5],[-1,2],[-9],[0,2]]
\end{curry}
It should be noted that \code{valuesOf} enumerates all values
of the given type completely and without duplicates.\footnote{%
In order to get an idea of the distribution of the generated test data,
CurryCheck also provides property combinators
\fcode{collect} and \fcode{classify} known from QuickCheck.}
Hence, if the set of possible input values is finite,
it is ensured that all of them are tested if sufficiently many tests
are performed. In this case, the property is also verified
(where QuickCheck or PrologCheck does not give such guarantees).
For instance, consider the De Morgan law from Boolean algebra:
\begin{curry}
negOr b1 b2 = not (b1 || b2) -=- not b1 && not b2
\end{curry}
This property is proved by CurryCheck after four tests
with all possible input values, and the output of CurryCheck
indicates that the testing was exhaustive:
\begin{curry}
negOr (module BoolTest, line 4):
 Passed all available tests: 4 tests.
\end{curry}

%%%%%%%%%%%%%%%%%%%%%%%%%%%%%%%%%%%%%%%%%%%%%%%%%%%%%%%%%%%%%%%%%%%%%%%%%
\section{User-Defined Test Data}
\label{sec:gentestdata}

Due to the use of functional logic features to generate test data,
one can write properties not only on predefined data types
but also on user-defined data types.
For instance, consider the following definition
of general polymorphic trees:
\begin{curry}
data Tree a = Leaf a | Node [Tree a]
\end{curry}
We define operations to compute the leaves of a tree and mirroring a tree:
\begin{curry}
leaves (Leaf x)  = [x]
leaves (Node ts) = concatMap leaves ts$\listline$
mirror (Leaf x)  = Leaf x
mirror (Node ts) = Node (reverse (map mirror ts))
\end{curry}
The following properties should increase our confidence in the
correctness of the implementation:
\begin{curry}
doubleMirrorIsId t = mirror (mirror t) -=- t$\listline$
leavesOfMirrorAreReversed t = leaves t -=- reverse (leaves (mirror t))
\end{curry}
CurryCheck successfully tests these properties without providing
any further information about how to generate test data.
However, in some cases it might be desirable to define our own test data
since the generated structures are not appropriate for testing.
For instance, if we test algorithms working on balanced search trees,
we need correctly balanced search trees as test data.
As a naive approach, we can limit the tests to correct test inputs
by using conditional properties.
As a simple example, consider the following operation that adds all numbers
from 1 to a given limit:
\begin{curry}
sumUp n = if n==1 then 1 else n + sumUp (n-1)
\end{curry}
Since there is also a simple formula to compute this sum,
we can check it:
\begin{curry}
sumUpIsCorrect n = n>0 ==> sumUp n -=- n * (n+1) `div` 2
\end{curry}
Note that the condition is important since \code{sumUp}
diverges on non-positive numbers.
As a result, Curry\-Check tests this property by enumerating integers
and dropping tests with non-positive numbers.
While this works well, since CurryCheck performs a fairly good
distribution between positive and negative numbers,
this approach might have a serious drawback if the proportion
of correct test data is small.
In the case of balanced search trees,
there are many more unbalanced trees than balanced search trees.
This has the effect that CurryCheck generates many test data
and drops it so that it does not make much progress.
Actually, CurryCheck has an upper limit for dropping test data
in the conditional operator in order to avoid spending too much
work on generating unusable test data.
For instance, if we want to test the above property \code{revRevIsId}
on long input lists, we could define it as follows:
\begin{curry}
revRevIsIdLong :: [Int] -> Prop
revRevIsIdLong xs = length xs > 100 ==> rev (rev xs) -=- xs
\end{curry}
Since there are a huge number of integer lists with a length smaller than 100,
CurryCheck does not find any test case
(with a default limit of dropping at most 10,000 incorrect test data
values):
\begin{curry}
revRevIsIdLong (module Rev, line 13):
 Arguments exhausted after 0 test.
\end{curry}
This shows that the fully automatic generation of test data
is not always appropriate.
Therefore, CurryCheck provides some combinators to
explicitly define test data (more advanced enumeration combinators
in the context of Scala are discussed in \cite{KurajKuncakJackson15}).

To show the method for test data generation in more detail,
we have to review Curry's methods to encapsulate non-deterministic
computations. For this purpose, Curry defines the following structure
to represent the results of a non-deterministic computation
\cite{BrasselHanusHuch04JFLP}:
\begin{curry}
data SearchTree a = Value a | Fail | Or (SearchTree a) (SearchTree a)
\end{curry}
\code{(Value v)} and \code{Fail} represent a single value
or a failure (i.e., no value), respectively,
and \code{(Or t1 t2)} represents a non-deterministic choice
between two search trees \code{t1} and \code{t2}.
Furthermore, there is a primitive search operator
\begin{curry}
someSearchTree :: a -> SearchTree a
\end{curry}
which yields a search tree for an expression.
For instance, \code{someSearchTree (0?1)} evaluates to the
search tree
\begin{curry}
Or (Value 0) (Value 1)
\end{curry}
The expression
\begin{curry}[mathescape=false]
someSearchTree (id $## (_::[Bool]))
\end{curry} % $
(where \ccode{\$\#\#} is an infix application operator which
evaluates the right argument to ground normal form before applying
the left argument to it) yields an (infinite)
search tree of all Boolean lists:
\begin{curry}
(Or (Value []) (Or (Or (Or (Value [False]) ...) (Or ...)) ...))
\end{curry}
Basically, EasyCheck defines various strategies to traverse
such search trees (see \cite{ChristiansenFischer08FLOPS} for details)
in order to enumerate test data.
Hence, if we want to define our own test data,
we have to define an operation that generates a search tree
containing the test data in \code{Value} leaves.
Although this is not difficult for simple data types,
it could be demanding for polymorphic types where
generators for the polymorphic arguments must be weaved
with the generators for the main data structure.
To simplify this task, CurryCheck offers a family of
combinators \code{genCons$n$} where each combinator
takes an $n$-ary constructor function
and $n$ generators as arguments and produces a search tree
for this constructor and all combinations of generated arguments.
Hence, these combinators have the type
\begin{curry}
genCons$n$ :: ($a_1 \to \cdots \to a_n \to a$) -> SearchTree $a_1$ -> $\cdots$ -> SearchTree $a_n$
         -> SearchTree $a$

\end{curry}
Furthermore, there is an infix combinator \ccode{|||} to
combine two search trees.
For instance, consider the straightforward definition of
Peano numbers:
\begin{curry}
data Nat = Z | S Nat
\end{curry}
Then we can define a search tree generator for this type as follows:
\begin{curry}
genNat :: SearchTree Nat
genNat = genCons0 Z ||| genCons1 S genNat
\end{curry}
Similarly, we can define a search tree generator for polymorphic trees
which takes a search tree for the argument type as a parameter
(where \code{genList} denotes the corresponding generator for list values):
\begin{curry}
genTree :: SearchTree a -> SearchTree (Tree a)
genTree ta = genCons1 Leaf ta ||| genCons1 Node (genList (genTree ta))
\end{curry}
The explicit definition of value generators is reasonable
when only a subset of all values should be used for testing.
For instance, \code{sumUpIsCorrect} should be testest with
positive numbers only.
Hence, we define a generator for positive numbers:
\begin{curry}
genPos = genCons0 1 ||| genCons1 (+1) genPos
\end{curry}
Since these numbers are slowly increasing, i.e., the search tree
is actually degenerated to a list, we can also use
the following definition to obtain a more balanced search tree:
\begin{curry}
genPos = genCons0 1 ||| genCons1 (\n -> 2*(n+1)) genPos
                    ||| genCons1 (\n -> 2*n+1)   genPos
\end{curry}
In order to test properties with user-defined data,
CurryCheck provides the property combinator
\begin{curry}
forAll :: [a] -> (a -> Prop) -> Prop
\end{curry}
which is satisfied if the parameterized property given as the second
argument is satisfied for all values of the first argument list.
Since there is also a library operation
\begin{curry}
valuesOfSearchTree :: SearchTree a -> [a]
\end{curry}
(actually, the operation \code{valuesOf} introduced in
Sect.~\ref{sec:valuesOf} is defined via this operation) to enumerate
all values of a search tree, we can redefine the property
\code{sumUpIsCorrect} as follows:
\begin{curry}
sumUpIsCorrect = forAll (valuesOfSearchTree genPos)
                        (\n -> sumUp n -=- n*(n+1) `div` 2)
\end{curry}
Using this technique, we could also define finite tests
for potentially infinite structures, e.g., one can easily define
tree generators that generate all trees up to a particular depth.

Finally, we show the implementation of the
combinators to generate search trees.
The definition of \ccode{|||} and \code{genCons0} is straightforward:
\begin{curry}
x ||| y = Or x y$\listline$
genCons0 v = Value v
\end{curry}
To define the further combinators like \code{genCons1},
we have to replace in a given search tree (for the argument)
the \code{Value} nodes by new nodes where the constructor operation
is applied to the given value.
This task is done by the following auxiliary operation:\footnote{%
This operation is similar to the monadic bind operation
in Haskell's \code{MonadPlus}, but we use this definition
due to the lack of type classes in the current language definiton
of Curry.}
\begin{curry}
updateValues :: SearchTree a -> (a -> SearchTree b) -> SearchTree b
updateValues (Value a)  f = f a
updateValues Fail       f = Fail
updateValues (Or t1 t2) f = Or (updateValues t1 f) (updateValues t2 f)
\end{curry}
The definition of the remaining combinators is now easy
(we only show the first two ones):
\begin{curry}
genCons1 c gena = updateValues gena (\a -> Value (c a))$\listline$
genCons2 c gena1 gena2 =
  updateValues gena1 (\a1$\;$->$\;$updateValues gena2 (\a2$\;$->$\;$Value (c a1 a2)))
\end{curry}

%%%%%%%%%%%%%%%%%%%%%%%%%%%%%%%%%%%%%%%%%%%%%%%%%%%%%%%%%%%%%%%%%%%%%%%%%
\section{Contract and Specification Testing}
\label{sec:contracts}

As discussed in detail in \cite{AntoyHanus12PADL},
the distinctive features of Curry (e.g., non-deterministic operations,
demand-driven evaluation, functional patterns, set functions)
support writing high-level specifications
as well as efficient implementations for a given problem
in the same programming language.
When applying this idea, Curry can be used
as a wide-spectrum language \cite{BauerEtAl78} for software development.
If a specification or contract is provided for some function,
one can exploit this information to support
run-time assertion checking with these specifications and contracts.
As an additional use of this information,
CurryCheck automatically generates properties
to test the given specifications and contracts,
which is described in the following.

According to the notation proposed in \cite{AntoyHanus12PADL},
a \emph{specification}\index{specification}
for an operation $f$ is an operation \code{$f$'spec}
of the same type as $f$.
A \emph{contract}\index{constract} consists
of a pre- and a postcondition, where the precondition could be omitted.
When provided, a \emph{precondition}\index{precondition} for an operation $f$
of type $\tau \to \tau'$ is an operation
\begin{curry}
$f$'pre :: $\tau$ ->$~$Bool
\end{curry}
putting demands on allowed arguments, whereas
a \emph{postcondition}\index{postcondition} for $f$
is an operation
\begin{curry}
$f$'post :: $\tau$ ->$~\tau'$ ->$~$Bool
\end{curry}
which relates input and output values
(the generalization to operations with more than one argument
is straightforward).
A specification should precisely describe the meaning of an operation,
i.e., the declarative meaning of the specification and the implementation
of an operation should be equivalent.
In contrast, a contract is a partial specification, e.g.,
all results computed by the implementation should satisfy the
postcondition.

As a concrete example, consider the problem of sorting a list.
The specification defines a sorted version of a given list
as a permutation of the input which is sorted.
Exploiting the operations introduced in Sect.\ref{sec:properties},
we define the following specification for the operation \code{sort}:
\begin{curry}
sort'spec :: [Int] -> [Int]
sort'spec xs | ys == perm xs && sorted ys = ys  where ys free
\end{curry}
A postcondition, which is easier to check, states that the
input and output lists should have the same length:
\begin{curry}
sort'post :: [Int] -> [Int] -> Bool
sort'post xs ys = length xs == length ys
\end{curry}
To provide a concrete implementation, we implement the quicksort algorithm
as follows:
\begin{curry}
sort :: [Int] -> [Int]
sort []     = []
sort (x:xs) = sort (filter (<x) xs) ++ [x] ++ sort (filter (>x) xs)
\end{curry}
Note that specifications and contracts are optional.
However, if they are included in a module processed with CurryCheck,
CurryCheck automatically generates and checks properties that
relate the specification and contract to the implementation.
For instance, an implementation satisfies a specification
if both yield the same values, and a postcondition is satisfied
if each value computed for some input satisfies the postcondition
relation between input and output. For our example, CurryCheck generates
the following properties (if there are also
preconditions for some operation, these preconditions are used
to restrict the test cases via the condition operater \ccode{==>}):\footnote{%
The property ``{\fcode{always$\;x$}}'' is satisfied
if all values of $x$ are \fcode{True}.}
\begin{curry}
sortSatisfiesSpecification :: [Int] -> Prop
sortSatisfiesSpecification x = sort x <~> sort'spec x$\listline$
sortSatisfiesPostCondition :: [Int] -> Prop
sortSatisfiesPostCondition x = always (sort'post x (sort x))
\end{curry}
With CurryCheck, the framework of \cite{AntoyHanus12PADL}
becomes more useful since contracts are not only used
as run-time assertions in concrete computations,
but many possible computations are checked with various test data.
For instance, CurryCheck reports that the above implementation of
\code{sort} is incorrect for the example input \code{[1,1]}
(as the careful reader might have already noticed).
When reporting the error, the module and source code line number
of the erroneous operation is shown so that the programmer
can easily spot the problem.

Another kind of contracts taken into account by CurryCheck
are determinism annotations.
An operation that yields always identical results
(maybe multiple times) on identical argument values
can be annotated as ``deterministic'' by adding \code{DET} to the
result type of its type signature.
For instance, the following operation tests whether a list
represents a set, i.e., has no duplicate elements (the definition
exploits functional patterns \cite{AntoyHanus05LOPSTR} as
well as default rules \cite{AntoyHanus16PADL}):
\begin{curry}
isSet :: [a] ->DET Bool
isSet (_++[x]++_++[x]++_) = False
isSet'default _           = True
\end{curry}
The determinism annotation \ccode{$\to$DET} has the effect that 
at most one result is computed for a given input, e.g., a single value
\code{False} is returned for the call
\code{isSet$\;$[1,3,1,3,1]}, although the first rule can be applied
multiple times to this call. Thus, after computing a first value,
all attempts to compute further values are ignored.
In order to ensure that this does not destroy completeness,
i.e., it behaves like ``green cuts'' in Prolog,
such operations must be deterministic from a semantical point of view.
CurryCheck tests this property by generating a property
for each \code{DET}-annotated operation
that expresses that there is at most one value for each input.
For instance, for \code{isSet}, the \code{DET} annotation is removed
and the property
\begin{curry}
isSetIsDeterministic x1 = isSet x1 #< 2
\end{curry}
is added by CurryCheck, where \ccode{$e\;$\#<$\;n$} is satisfied
if the \emph{set} of all values of $e$ contains less than $n$ elements.

%%%%%%%%%%%%%%%%%%%%%%%%%%%%%%%%%%%%%%%%%%%%%%%%%%%%%%%%%%%%%%%%%%%%%%%%
\section{Combining Testing and Verification}
\label{sec:verify}

The objective of CurryCheck is to increase the confidence
in the reliability of Curry programs.
Testing with a lot of input data is one important step
but, in case of input data types with infinite values,
it can only show possible errors but not the absence of them.
In order to support the latter,
CurryCheck has also some (preliminary) support
to include the verification of program properties.
For this purpose, a programmer might prove properties
stated in a source program.
Since there are many different possibilities to prove such properties,
ranging from manual proofs to interactive proof assistants
and fully automatic provers, CurryCheck does not enforce
a particular proof technique.
Instead, CurryCheck trusts the programmer and uses a naming
convention for files containing proofs:
if there is a file with name \code{proof-$t$.*},
CurryCheck assumes that this file contains
a valid proof for property $t$.
For instance, the following property states that
sorting a list does not change its length:
\begin{curry}
sortlength xs = length (sort xs) <~> length xs
\end{curry}
If there is a file \code{proof-sortlength.agda},
containing a proof for the above property
(\cite{AntoyHanusLibby16} addresses techniques how to prove such
properties in the dependently typed language Agda),
CurryCheck considers this property as valid and does not check it.
Moreover, it uses it to simplify other properties to be tested.
In our case, the property \code{sortSatisfiesPostCondition}
of the previous section can be simplified to
\code{always$\;$True} so that it does not need to be tested.
Similarly, a determinism annotation for operation $f$ is not tested
if there is a proof file \code{$f$IsDeterministic.*}.

Since program verification is a notoriously difficult task,
a mixture of different techniques is required.
For instance, \cite{JohanssonEtAl14} discusses techniques
to use the Isabelle/HOL proof assistant to verify purely functional
properties inspired by QuickCheck.
\cite{AntoyHanusLibby16} describes a method to prove
non-deterministic computations by translating Curry programs
into Agda programs. Since these proofs can be verified by the Agda compiler,
CurryCheck can test the validity of a given proof file by simply
invoking the Agda compiler.
Some purely functional properties can be proved
in a fully automatic way. For instance, the properties
\begin{curry}
concLength xs = length (xs ++ ys) -=- length xs + length ys
revLength  xs = length (rev xs) -=- length xs
\end{curry}
can be proved by the SMT solver Alt-Ergo.
To support the use of such solvers,
we have started the development of tools to automatically
translate Curry programs into the syntax of Agda and other
proof systems. We omit more details since this is outside the
scope of this paper.

%%%%%%%%%%%%%%%%%%%%%%%%%%%%%%%%%%%%%%%%%%%%%%%%%%%%%%%%%%%%%%%%%%%%%%%%
\section{Practical Experience}
\label{sec:impl}

The implementation of CurryCheck is available with the
(Prolog-based) Curry implementation PAKCS \cite{Hanus16PAKCS} (since version
1.14.0) and the
(Haskell-based) Curry implementation KiCS2 \cite{BrasselHanusPeemoellerReck11}
(since version 0.5.0).
The implementation exploits meta-programming features
available in these systems to parse programs and transform
them into new programs as described in the previous sections.

Although we could show in this paper only simple examples,
we would like to remark that CurryCheck is successfully applied
in a larger context.
CurryCheck is regularly used for automatic regression testing
during continuous integration and nightly builds of PAKCS and KiCS2.
Currently, approximately 500 properties (the number is continuously growing)
are regularly used to test the libraries and tools of these systems.
Our practical experience is quite promising.
After the development and use of CurryCheck,
we found a bug in the implementation of the prelude operations
\code{quot} and \code{rem} w.r.t.\ negative numbers and free variables
which was undetected for a couple of years.
Although the bug was easy to fix, the definition of a general property
relating both operations and testing it
with all smaller values was essential for its detection.

The run time of CurryCheck mainly depends on the specific
properties to be tested.
The initial translation phase, which extracts properties,
contracts, and specifications from a given module and transforms
them into executable tests, is a straightforward compilation
process.
The run time of the subsequent test execution phase
depends on the number of test cases and the time needed to evaluate
each property.
The functional logic programming technique to generate
test data described in Sect.~\ref{sec:testproperties}
(i.e,. collecting all non-deterministic results of evaluating
a logic variable)
is reasonable in practice.
For instance, KiCS2 needs 0.6 seconds to test a trivial property
on a list of integers with 10,000 test cases
computed by the randomized level diagonalization
strategy described in \cite{ChristiansenFischer08FLOPS}
(on a Linux machine with Intel Core i7-4790/3.60Ghz and 8GiB of memory).

%%%%%%%%%%%%%%%%%%%%%%%%%%%%%%%%%%%%%%%%%%%%%%%%%%%%%%%%%%%%%%%%%%%%%%%%
\section{Related Work}
\label{sec:related}

Since testing is an important part of the software development process,
there is a vast literature on this topic. In the following,
we compare our approach to testing, in particular, property-based testing,
in declarative languages. We already mentioned
QuickCheck \cite{ClaessenHughes00} which was influential
in this area and followed by other property-testing systems
for functional languages, like
GAST \cite{KoopmanAlimarineTretmansPlasmeijer03}
or SmallCheck \cite{RuncimanNaylorLindblad08}.
The same idea has also been transferred to other languages
like PropEr \cite{PapadakisSagonas11} for Erlang and
PrologCheck \cite{AmaralFloridoSantosCosta14} for Prolog.
In contrast to CurryCheck, most of these systems (except for SmallCheck)
are based on randomly
generating test data so that they do not provide guarantees for a complete
enumeration if the sets of input values are finite, i.e., they cannot verify
properties.
PropEr also supports contract checking but these function contracts are limited
to type specifications.
PrologCheck could also check operational aspects likes modes
or multiplicity of answers, whereas our properties
are oriented towards declarative aspects, i.e., the input/output
relation of values.

Closely related to CurryCheck is EasyCheck \cite{ChristiansenFischer08FLOPS}
since it can be seen as our back end.
EasyCheck is the only property-based test tool
covering functional and logic aspects but it is more limited than CurryCheck.
EasyCheck does not support polymorphic properties, I/O properties, or
combinators for user-defined generation of test data.
This has been added in CurryCheck together with a full automation
of the test process in order to obtain an easily usable tool.
Moreover, CurryCheck expands the use of automatic
testing by using it for contract and specification checking,
where functional logic programming has been shown to be an appropriate
framework \cite{AntoyHanus12PADL},
and combines it with static verification techniques.

%%%%%%%%%%%%%%%%%%%%%%%%%%%%%%%%%%%%%%%%%%%%%%%%%%%%%%%%%%%%%%%%%%%%%
\section{Conclusion}
\label{sec:conclusion}

We have presented CurryCheck, the first fully automatic tool
to test functional as well as non-deterministic properties
of Curry programs. CurryCheck supports unit tests and tests of
I/O operations with fixed inputs as well as property tests
which are parameterized over some arguments.
In the latter case, they are executed with test inputs
which are systematically generated for the given argument types.
Moreover, CurryCheck also supports specification and contract testing
if such information is present in the source program.

To simplify and, thus, enhance the use property testing,
properties can be interspersed in the source program
and are automatically extracted by CurryCheck.
Hence, CurryCheck supports test-driven program development methods
like extreme programming.
Properties are not only useful to obtain more reliable programs,
but they can also be used by automated documentation tools
to describe the intended meaning of operations,
a feature which has been recently added to the
CurryDoc \cite{Hanus02WFLP} documentation tool.\footnote{%
See \url{www.informatik.uni-kiel.de/~pakcs/lib/Combinatorial.html}
for an example.}

For future work we plan to extend the functionality
of CurryCheck (the current version does not support
the generation of floating point numbers and functional values).
Furthermore, we intend to integrate into CurryCheck
more features that can help to improve the reliability of the
source code, like abstract interpretation to approximate
specific run-time properties
\cite{FaehndrichLogozzo11,StulovaMoralesHermenegildo16},
or program covering to show
whether the test data was sufficient to reach all parts
of a source program.

%%%%%%%%%%%%%%%%%%%%%%%%%%%%%%%%%%%%%%%%%%%%%%%%%%%%%%%%%%%%%%%%%%%%%
\paragraph{Acknowledgements.}
The author is grateful to Jan-Patrick Baye
for implementing an initial version of CurryCheck.

%\bibliographystyle{plain}
%\bibliography{mh}

\end{document}